\documentclass[12pt]{article}

\usepackage{epsf}
\usepackage{amsmath,amssymb}
\usepackage{latexsym}
\usepackage{graphicx,color}
\usepackage{wrapfig}
\usepackage{latexsym}
\usepackage{graphics}

\usepackage{color}
\usepackage{delarray,color}
\usepackage{fancybox}  
\newcommand {\beq}{\begin{eqnarray}}
\newcommand {\eeq}{\end{eqnarray}}

\newcommand{\de}{\partial}
\newcommand{\be}{\begin{equation}}
\newcommand{\ba}{\begin{eqnarray}}
\newcommand{\ea}{\end{eqnarray}}
\newcommand{\ee}{\end{equation}}

\newcommand{\beqa}{\begin{eqnarray}}
\newcommand{\eeqa}{\end{eqnarray}}
\newcommand{\CR}{\nonumber \\}

\newcommand{\unit}{\hbox to 3.8pt{\hskip1.3pt \vrule height 7.4pt
    width .4pt \hskip.7pt \vrule height 7.85pt width .4pt \kern-2.4pt
    \hrulefill \kern-3pt \raise 3.7pt\hbox{\char'40}}}

\def\matt[#1,#2,#3,#4]{\left(%
\begin{array}{cc} #1 & #2 \\ #3 & #4 \end{array} \right)}

\usepackage{tabularx}

\catcode`\@=11
\@addtoreset{equation}{section}

\catcode`@=12
\relax

\begin{document}

\baselineskip 0.7cm

\begin{titlepage}

\setcounter{page}{0}

\renewcommand{\thefootnote}{\fnsymbol{footnote}}

\begin{flushright}
YITP-10-98\\
IHES/P/10/45
\end{flushright}

\vskip 1.35cm

\begin{center}
{\Large \bf 
On Effective Action of Multiple M5-branes and ABJM Action
}

\vskip 1.2cm 

{\normalsize
Seiji Terashima$^1$\footnote{terasima(at)yukawa.kyoto-u.ac.jp} and Futoshi
Yagi$^2$\footnote{fyagi(at)ihes.fr}
}

\vskip 0.8cm

{ \it
$^1$Yukawa Institute for Theoretical Physics, Kyoto University, Kyoto 606-8502,
Japan \\
$^2$Institut des Hautes \'Etudes Scientifiques, Bures-sur-Yvette, 91440, France
}

\end{center}

\vspace{12mm}

\centerline{{\bf Abstract}}

We calculate the fluctuations from the classical 
multiple M5-brane solution
of ABJM action which we found in the previous paper.
We obtain D4-brane-like action but the gauge coupling constant depends
on the spacetime coordinate.
This is consistent 
with the expected properties of M5-brane action, 
although 
we will need 
to take into account the monopole operators 
in order to fully understand M5-branes.
We also see that the Nambu-Poisson bracket is hidden in the solution.

\end{titlepage}
\newpage

\tableofcontents

\section{Introduction}\label{Intro}

Understanding the dynamics of M5-branes is one of the most important 
problems in M-theory or string theory. 
For the D-branes, the low energy effective actions 
were found and they are essentially the Yang-Mills action. 
However, the effective action for the
multiple M5-branes is not known yet,
although single M5-brane action is known \cite{Perry, Schwarz, PST}.
Actually, from the AdS/CFT correspondence, the degree of freedom
of the $N$ M5-branes will be proportional to $N^3$,
which can not be realized by the Yang-Mills theory naively.
Thus, there should be interesting physics which is not yet 
known behind it.

Recently, the effective action of multiple M2-branes 
on $\mathbb{C}^4/\mathbb{Z}_k$
was suggested by \cite{ABJM} (we will call the action as ABJM action), 
following \cite{BL1,BL2,BL3,G}.
Because the D4-brane action is constructed from the D2-brane action
by the Matrix theory like construction \cite{BSS, BFSS}
using the non-commutative space \cite{CoDoSc}-\cite{Se},
we expect that this ABJM action will be useful to
study M5-branes. 
More concretely, it is known that the effective action of $N$ D2-branes,
which is three dimensional supersymmetric Yang-Mills theory, 
in a large $N$ limit has a
classical solution which correspond to 
D4-branes with a constant magnetic flux.
This is an D2-D4 bound state and the flux represents the non-zero 
D2-brane charge.
When we consider small fluctuation around 
this classical solution, we obtain the effective 
D4-brane action,
which is five dimensional supersymmetric Yang-Mills theory.
We would like to uplift this situation to M-theory
in order to obtain an M5-brane action from
an M2-brane action, which we expect to be the ABJM action%
\footnote{In \cite{HM, HIMS}, 
a single M5-brane is constructed
from BLG model by using Nambu-Poisson bracket as the three-algebra.
However, 
multiple M5-branes has not been obtained.}.

Several classical solutions of M2-M5 bound state
have been found in the ABJM action.
One of them is the M2-branes ending on M5-branes.
Such a classical solution was first studied in \cite{Terashima}
and a closely related solution was also found in the mass deformed ABJM action
\cite{GRVV}%
\footnote{For still another closely related work, see also \cite{HL}.}.
Although both of them are expected to form a fuzzy $S^3/\mathbb{Z}_k$,
they actually show a fuzzy $S^2$ in the naive classical analysis.
The fluctuation around this classical solution 
was also calculated \cite{Nastase1} 
and shown that it actually reduces to D4-brane action%
\footnote{It is discussed in \cite{Nastase1} that
we have to take the limit $k \to \infty$
in order that this analysis is reliable.
In this limit, M5-branes reduce to D4-branes.}.
Indeed, the classical solution \cite{GRVV} of the mass deformed ABJM model 
is shown to be exactly equivalent to the usual fuzzy 
$S^2$ solution corresponding to the D2-D4 bound state
constructed with the adjoint scalar field at the classical level%
\footnote{Since this analysis is purely classical, the contribution
from the monopole operator, which we mention later, is not included.}
\cite{Nastase1}.

In the previous paper \cite{previous}, 
we found another classical solution of M5-branes.
This solution is an uplift of the flat D4-brane solution with
a constant magnetic flux,
which is constructed from infinitely many D2-branes%
\footnote{Since this D4-brane solution exists in the strict large $N$ limit
contrary to the case of the fuzzy sphere which has a finite volume,
so does our M5-brane solution.
In this paper, our discussion is limited to the case where
the number of the M2-branes is strictly infinite.}
satisfying $[X^1, X^2] = const.$ where $X^1$ and $X^2$ 
are the adjoint scalar fields corresponding to the transverse 
direction of the D2-branes like \cite{BSS}.
The three-algebra structure was also found in this solution. 
In this paper, we expand ABJM action around our classical solution.
We obtain D4-brane like action,
which contains only the zero modes of the $S^1$ direction
on which the $\mathbb{Z}_k$ of the $\mathbb{C}^4/\mathbb{Z}_k$ acts.
This is because the non-zero modes should 
have the vortex (or monopole) charge through the Chern-Simons term.
In order to include them, we should take into account the monopole operators 
\cite{ABJM}. 

However, it would be remarkable that
the gauge coupling constant of our D4-brane like action
depends on the spacetime coordinate.
We would like to stress that such an action is not obtained from the D2-branes.
This dependence reflects the geometry $\mathbb{C}^4/\mathbb{Z}_k$ 
in which the radius of the $S^1$ increases 
as we go away from the orbifold fixed point.
In this sense, our action includes the information of M5-brane,
which is not included in the D4-brane constructed from D2-branes.
Although our result 
may not include all the low energy dynamics of the M5-branes,
we hope it will still be helpful for the understanding of the M5-branes.

\section{ABJM action and the M5-brane solution}

In this section, we review ABJM action
and our classical M5-brane solution.
Then, we discuss that the structure of
Nambu-Poisson bracket is hidden in our classical solution.

\subsection{ABJM action}

ABJM action is a three dimensional $N=6$
supersymmetric Chern-Simons theory,
whose gauge group is $U(N)_1 \times U(N)_2$.
Its matter contents are 
gauge fields $A^{(1)}_{\mu}$, $A^{(2)}_{\mu}$
of each gauge group,
four complex bi-fundamental scalar fields $Y^A$ ($A=1,2,3,4$) 
and their fermionic superpartners.
The bosonic part of the ABJM action is given by
\begin{eqnarray}
L= \frac{k}{4\pi} \varepsilon^{\mu\nu\rho}
\mathrm{Tr} \left( A^{(1)}_{\mu} \partial_{\nu} A^{(1)}_{\lambda} 
+ \frac{2i}{3} A^{(1)}_{\mu} A^{(1)}_{\nu} A^{(1)}_{\lambda} 
- A^{(2)}_{\mu} \partial_{\nu} A^{(2)}_{\lambda} 
- \frac{2i}{3} A^{(2)}_{\mu} A^{(2)}_{\nu} A^{(2)}_{\lambda} \right) \nonumber \\
- \mathrm{Tr} \left[ ( \tilde{D}_{\mu} Y_A )^{\dagger} \tilde{D}^{\mu} Y^A \right]
- V_{\rm bos}.
\end{eqnarray}
The bosonic potential $V_{\rm bos}$ is given by 
\begin{eqnarray}
V_{\rm bos} = 
- \frac{4\pi^2}{3k^2} \mathrm{Tr} 
\left( \{ \Upsilon^{AB}{}_{C} ,  \Upsilon_{AB}{}^{C}\} \right),
\end{eqnarray}
where
\begin{eqnarray}
\Upsilon^{AB}{}_{C} = 
\left[ Y^A, Y^B ; Y_C \right]
- \frac{1}{2} \delta^{A}_{C} \left[ Y^D, Y^B ; Y_D \right]
+ \frac{1}{2} \delta^{B}_{C} \left[ Y^D, Y^A ; Y_D \right],
\end{eqnarray}
and the bracket $\{ \,\, ,  \,\, \}$ is the anti-commutator.
Here, the three-bracket is defined as
\begin{eqnarray}
[X,Y;Z] = XZY - YZX.
\end{eqnarray}
The scalar fields with lower indices are given by
\begin{eqnarray}
Y_A = (Y^A)^{\dagger}.
\end{eqnarray} 
By using the definition above,
the bosonic potential can be explicitly written 
in terms of three-bracket as
\begin{eqnarray}
V_{\rm bos} 
&=&  
- \frac{4\pi^2}{3k^2} \mathrm{Tr} \Bigl(
\left\{ \left[ Y^A, Y^B ; Y_C \right] , 
\left[ Y_A, Y_B ; Y^C \right] \right\} 
\cr
&&  \left. \qquad \qquad \quad
- \frac{1}{2} \left\{ \left[ Y^A, Y^C ; Y_A \right] ,
\left[ Y_B, Y_C ; Y^B \right] \right\} \right).
\label{bos}
\end{eqnarray}

For later convenience,
we define new basis of the gauge fields as
\begin{eqnarray}
A_{\mu} \equiv \frac{1}{2} \left( A^{(1)}_{\mu} + A^{(2)}_{\mu} \right),
\qquad 
B_{\mu} \equiv \frac{1}{2} \left( A^{(1)}_{\mu} - A^{(2)}_{\mu} \right).
\label{gauge_A}
\end{eqnarray}
Rewriting the Chern-Simons term with these new basis, we obtain
\begin{eqnarray}
L_{\rm CS} 
= \frac{k}{2\pi} \varepsilon^{\mu\nu\rho}
\mathrm{Tr} \left( B_{\mu} F_{\nu\lambda}
+ \frac{2i}{3} B_{\mu} B_{\nu} B_{\lambda}
\right),
\end{eqnarray}
where we put
\begin{eqnarray}
F_{\nu\lambda} = \partial_{\nu} A_{\lambda} 
- \partial_{\nu} A_{\lambda} + i [A_{\nu}, A_{\lambda}] .
\end{eqnarray}
The covariant derivatives for the bi-fundamental fields
$Y^A$ are also rewritten
in terms (\ref{gauge_A}) as
\begin{eqnarray}
\tilde{D}_{\mu} Y^A 
= D_{\mu} Y^A + i\{ B_{\mu}, Y^A \}
\label{cov}
\end{eqnarray}
where we put
\begin{eqnarray}
D_{\mu} Y^A = \partial_{\mu} Y^A + i [A_{\mu},Y^A].
\end{eqnarray}
The bosonic part of the ABJM action is then rewritten as
\begin{eqnarray}
L &=& \frac{k}{2\pi} \varepsilon^{\mu\nu\rho}
\mathrm{Tr} \left( B_{\mu} F_{\nu\lambda}
+ \frac{2i}{3} B_{\mu} B_{\nu} B_{\lambda}
\right)
\cr
&&- {\rm Tr} \left( D_{\mu} Y_A + i\{ B_{\mu}, Y_A \} \right)^{\dagger}
\left( D^{\mu} Y^A + i\{ B^{\mu}, Y^A \} \right)
- V_{\rm bos} .
\label{ABJMaction2}
\end{eqnarray} 

The moduli space of this theory is $(\mathbb{C}^4/ \mathbb{Z}_k)^N/S_{N}$,
where $\mathbb{Z}_k$ simultaneously rotate the phase of all 
the complex scalar fields $Y^i$ by $2\pi/k$.
Thus, the ABJM model is supposed to describe the $N$ M2-branes
probing $\mathbb{C}^4/ \mathbb{Z}_k$.
If we take the limit $k \to \infty$ and 
look far away from the orbifold fixed point at the same time,
the local geometry of $\mathbb{C}^4/ \mathbb{Z}_k$ becomes cylinder.
Thus, the M2-branes can be regarded as 
D2-branes probing $\mathbb{R}^7$ in this limit.
Indeed, when we give a vacuum expectation value $v$ to one of 
the scalars $Y^i$ and expand around that vacuum,
and consider the following limit;
\begin{eqnarray}
k,v \to \infty  \,\, \mathrm{with} \qquad \frac{k^2}{32 \pi^2 v^2}
=\frac{1}{4 g_{YM}^2} \,\, \mathrm{fixed}, 
\label{limit1}
\end{eqnarray}
we obtain the D2-branes low energy effective action, i.e. 
the super Yang-Mills theory \cite{ABJM, Mukhi}.
Due to the Higgs mechanism, 
the field $B_{\mu}$ becomes massive and integrated out
while $A_{\mu}$ remains as a gauge field on the D2-branes.
We denote the limit (\ref{limit1}) as the scaling limit in this paper.

\subsection{M5-brane solution}

In the previous paper \cite{previous},
we showed the existence and uniqueness, up to some trivial ambiguities,
of the solution of the following form
of the equations of motion 
for $U(N) \times U(N)$ ABJM action with $N \rightarrow \infty$:
\begin{eqnarray}
&&Y^1 = Y_1 = 1_{n \times n} \otimes r(\hat{x},\hat{y}) , \quad
Y^2 = Y_2 = 1_{n \times n} \otimes r'(\hat{x},\hat{y}), \nonumber \\
&&Y^3 = 0 ,\quad Y^4 = 0, \nonumber \\
&&A^{(1)}_{\mu} = A^{(2)}_{\mu} = 0,
\label{ansatz1}
\end{eqnarray}
where 
\begin{eqnarray}
r(\hat{x},\hat{y}) = v + \hat{x} + {\cal O} (v^{-1}), \quad
r'(\hat{x},\hat{y}) = \hat{y}, \quad
[\hat{x}, \hat{y}] = i\Theta,
\label{ansatz2}
\end{eqnarray}
and we regard that $\hat{x}$ and $\hat{y}$ are
infinite dimensional irreducible hermitian matrices.
This solution is constructed so that 
it reduces  in the scaling limit 
to the solution representing the $n$ D4-branes 
in the action of infinitely many 
D2-branes.
Thus, we interpret this classical solution as 
a solution representing $n$ M5-branes.
We found the explicit form of $r(\hat{x},\hat{y})$
pertubatively in ${\cal O} (v^{-9})$ \cite{previous}.

Interestingly, in the commutative limit $\Theta \to 0$\footnote{
This limit will correspond to 
the large background magnetic field in the D4-brane picture.
The reason why the vanishing magnetic field limit
 does not correspond
to $\Theta \rightarrow 0$ limit is that
we use matrix model like construction of the solution
\cite{Se}.
},
we can replace the commutator by Poisson bracket
and found the solution:
\begin{eqnarray}
[r , r']_P = \frac{i\Theta v}{\sqrt{r^2+{r'}^2}},
\label{M5_solution}
\end{eqnarray}
where $[ \,\, , \,\, ]_P$ represents the Poisson bracket
and $v$ is a constant\footnote{
The anti-bracket $\{ f, g \}$
will be approximated as $2 f g$ using the star-product formalism
in the limit
and the equations of motion become
\beqa
0 &\simeq& ((Y^1)^2)+(Y^2)^2))[Y^2,[Y^1,Y^2]_P ]_P 
- Y^1 ([Y^1,Y^2]_P)^2 \CR
0 &\simeq& ((Y^1)^2)+(Y^2)^2))[Y^1,[Y^1,Y^2]_P ]_P 
+ Y^2 ([Y^1,Y^2]_P)^2.
\eeqa
The general solution of these equations is indeed (\ref{M5_solution}).
}.
Here the coordinates become commutative in the limit 
and we denote them 
as $r$, and $r'$. 
More explicitly, from the above equation with the ansatz, 
the function $r(x,y)$ is determined by
\begin{eqnarray}
\frac{\partial r(x,y)}{\partial x} = \frac{v}{\sqrt{r^2 + y^2}},
\label{solr}
\end{eqnarray}
with 
\beq
[ x , y ]_P=i \Theta.
\label{poi}
\eeq
Note that 
the M5-branes span $\{ r,r',\theta\}$
where $Y^1=r e^{i \theta}$ and $Y^2=r' e^{i \theta}$.
The $\theta$-direction corresponds to the $S^1$ direction
which is mentioned in section \ref{Intro}.
The induced metric on the M5-branes 
is given by
\beq
ds^2 = ds_{(3)}{}^2 + dr^2 + dr'^2 + (r^2+r'^2) d\theta^2,
\label{metric}
\eeq
where $ds_{(3)}{}^2$ is 
the flat metric of the 1+2 dimensional Minkowski space-time
and then,
the Nambu-Poisson bracket naturally defined 
as $[x,y,\theta]_{NP} \sim 
\frac{1}{\sqrt{{\rm det} g_{ij}}} 
\epsilon^{ijk} \de_i x \de_j y \de_k \theta$ 
where $i,j,k =\{ r,y,\theta\}$
on this space 
is constant.
However, if we naively define 
a Poisson bracket on the dimensionally reduced 
space with $ds^2=dr^2+{dr'}^2$ 
as $[x,y]'_P \sim \frac{1}{\sqrt{{\rm det} g_{ij}}} 
\epsilon^{ij} \de_i x \de_j y$ 
where $i,j =\{ r,y \}$, then it is not constant
by (\ref{solr})
and is different from (\ref{poi}).

Thus, the above Poisson bracket will have to be regarded 
as the Nambu-Poisson bracket with one variable always chosen
to 
$\theta$,
\beq
[f(r,r') , g(r,r')]_{P} = [f(r,r') , g(r,r'), \theta]_{NP},
\eeq
where
\beq
[r , r', \theta]_{NP} = \frac{i\Theta v}{\sqrt{r^2+{r'}^2}},
\eeq
with 
\beq
[ x , y, \theta ]_{NP}=i \Theta.
\eeq
It can be written as 
\beq
[r , r', \theta]_{NP} = {i\Theta v} \sqrt{ {\rm det} \, g_{ij}}^{-1},
\eeq
where $g_{ij}$ is the metric on the space spanned by 
$\{ r,r',\theta \}$.

In the following section, we consider fluctuations
from this classical solution\footnote{
It is more appropriate to express the solution
as $[r , r', \theta]_{NP} = \frac{C}{\sqrt{r^2+{r'}^2}}$, where
$C$ is a constant which represents a strength of the background 
3-form field strength. If we focus on the solution near $r=v$,
then the non-commutative parameter $\Theta$ of the 
effective D5-brane action is $v$-dependent as $\Theta=C/v$.} 
in the commutative limit 
$\Theta \rightarrow 0$.

\section{The M5-action from the ABJM action}

In this section, we will expand the ABJM action around 
our classical solution (\ref{M5_solution})
 and find the action for the multiple M5-branes,
although we can keep only the zero-mode for the $\theta$ direction in the action.
We will see the action has a form of the Yang-Mills
action with a space-time dependent gauge coupling constant.

\subsection{Expansion of the bosonic potential}

Before expanding the ABJM action around our classical solution,
we rewrite the bosonic potential (\ref{bos}) for later convenience.
Since we will later use the classical solution (\ref{M5_solution}), 
which is valid in the commutative limit,
we expand the potential term 
by the number of the commutators.
As will be explained later,
we will take the fluctuations such 
that one commutator in the potential is ${\cal O} (\Theta)$.

First, we rewrite the three-bracket
by using commutator and anti-commutator as
\begin{eqnarray}
\left[ Y^A, Y^B ; Y_C \right] 
= \frac{1}{2} \left( 
  \left\{ [Y^A, Y^B ] , Y_C \right\}
+ \left\{ [Y^A, Y_C] , Y^B \right\}
- \left\{ [Y^B, Y_C ] , Y^A \right\}
\right),
\label{3_comm}
\end{eqnarray}
which is shown by using the graded Jacobi identity
\begin{eqnarray}
[\{A,B\},C] + \{[C,A],B\}-\{[B,C],A\} = 0.
\label{Graded_Jacobi}
\end{eqnarray}
Because the three-bracket can be represented as 
(\ref{3_comm}), the leading terms are terms with two commutators.
By substituting (\ref{3_comm}) into (\ref{bos}),
we find that the potential is given by 
\begin{eqnarray}
V_{\rm bos} &=& 
- \frac{4\pi^2}{3k^2} \mathrm{Tr} \left(
\left\{ \left[ Y^A, Y^B ; Y_C \right] , 
\left[ Y_A, Y_B ; Y^C \right] \right\} 
 - \frac{1}{2} \left\{ \left[ Y^A, Y^C ; Y_A \right] ,
\left[ Y_B, Y_C ; Y^B \right] \right\} \right) \CR
& =& -\frac{4\pi^2}{3k^2} \mathrm{Tr} \left(
|Y|^2 (-2 |[Y^A,Y^B]|^2 - 4 |[Y^A,Y_C]|^2 - |[Y^A,Y_A]|^2 )
\right. \CR 
& & \left. +5 |[Y^B,Y_A] Y^A|^2 +| [Y^B,Y^A] Y_A| 
\right. \CR 
& &   + \left. 3\left( [Y^A,Y^B]Y_B [Y_A,Y^C] Y_C+
[Y_A,Y_B]Y^B [Y^A,Y_C] Y^C \right) 
\left.  - 2 Y_c [Y^C,Y_A] Y^A [Y^B, Y_B] \right) 
\right. \CR 
& &  + {\cal O} ([\, ,]^3) 
\CR
&=& - \frac{4\pi^2}{3k^2} {\rm Tr}
(3A + B + 3C + 2D)
+ {\cal O} ([\, ,]^3) ,
\label{bos2}
\end{eqnarray}
where we have defined
\begin{eqnarray}
&&A = - |Y|^2 (|[Y^A,Y^B]|^2 + |[Y^A,Y_C]|^2) ,
\CR
&&B =|Y|^2 ( |[Y^A,Y^B]|^2 -  |[Y^A,Y_B]|^2 - |[Y^A,Y_A]|^2 ) ,
\CR
&&C = |[Y^B,Y_A] Y^A|^2 + |[Y^B,Y^A] Y_A|^2 + 
[Y^A,Y^B]Y_B [Y_A,Y^C] Y_C + [Y_A,Y_B]Y^B [Y^A,Y_C] Y^C ,
\CR
&&D = |[Y^B,Y_A] Y^A|^2 - |[Y^B,Y^A] Y_A|^2 - Y_C [Y^C,Y_A] Y^A [Y^B, Y_B].
\label{defs}
\end{eqnarray}

By using the identity
\begin{eqnarray}
[X,Y]Z = [XZ,Y] - X[Z,Y],
\label{id_xyz}
\end{eqnarray}
we can rewrite $B$ defined in (\ref{defs}) as
\begin{eqnarray}
B &\simeq& 
  |Y|^2 [Y^B , Y^A] [Y_A,Y_B] 
- |Y|^2 [Y^A , Y_B] [Y^B,Y_A]
+ |Y|^2 [Y^A , Y_A] [Y^B,Y_B]
\CR
&\simeq& 
  [Y^B, Y^A |Y|^2] [Y_A,Y_B] 
- [Y^A |Y|^2, Y_B] [Y^B,Y_A]
+ [Y^A |Y|^2, Y_A] [Y^B,Y_B]
\CR
&&- Y^A [Y^B, |Y|^2][Y_A, Y_B]
+Y^A [|Y|^2,Y_B] [Y^B,Y_A]
-Y^A [|Y|^2,Y_A][Y^B,Y_B]
\CR
&\simeq& 
[ Y^B [Y_A,Y_B] , Y^A|Y|^2]
- [Y^B[Y^A|Y|^2,Y_B] , Y_A]  
+ [Y^B[Y^A|Y|^2,Y_A] , Y_B]  
\CR
&&+ Y^B \left( 
- [[Y_A,Y_B] , Y^A|Y|^2]  
+ [[Y^A|Y|^2,Y_B] , Y_A]  
- [[Y^A|Y|^2,Y_A] , Y_B]  
\right)
\CR
&&+Y^A [|Y|^2, Y^B][Y_A, Y_B]+Y^A [|Y|^2,Y_B] [Y^B,Y_A]
+Y^A [|Y|^2,Y_A][Y_B,Y^B] .
\label{B0}
\end{eqnarray}
Each term in the first line after the last equality
is a commutator as a total and 
its trace vanishes, which we write 
``total div.'' in the following.
The second line identically vanishes due to 
the Jacobi identity.
The remaining part is the last line
and can be computed as
\begin{eqnarray}
B &\simeq& 
Y^A Y^C [Y_C,Y^B][Y_A,Y_B] + Y^A Y_C [Y^C,Y^B][Y_A,Y_B] \CR
&&
-Y^A Y^C [Y_A,Y^B][Y_C,Y_B] +Y^A Y_C [Y^C,Y_B][Y^B,Y_A] \CR
&&
+Y^A Y^C [Y_C,Y_A][Y_B,Y^B] +Y^A Y_C [Y^C,Y_A][Y_B,Y^B]
+ {\rm total \,div.}\CR
&=& D + {\rm total \,div.},
\label{B0_2}
\end{eqnarray}
where in the final line we used the symmetry between $A$ and $C$ indices.
Thus, we find that the potential (\ref{bos2}) simplifies as
\begin{eqnarray}
V_{\rm bos} = 
- \frac{4\pi^2}{k^2} {\rm Tr}(A + C + D)
+ {\rm total \,div.}
+ {\cal O} ([\, ,]^3) ,
\label{bos3}
\end{eqnarray}
which gives the potential up to two commutators.

For later convenience, we decompose the complex scalar fields $Y^A$ 
into the real part and the imaginary part as 
\beq
Y^A=p^A+i q^A,
\label{real_im}
\eeq
where $p,q$ are Hermite matrices.
By substituting this decomposition into $A$ in (\ref{defs}), we obtain
\beqa
A = 2 \left((p^A)^2 + (q^A)^2) ([p^B,p^C]^2 +2 [p^B,q^C]^2 +
[q^B,q^C]^2 \right) + {\cal O} ([\, ,]^3) ,
\label{A}
\eeqa
which is $SO(8)$ invariant.
It is also straightforward to show that
$C$ in (\ref{defs}) is given by
\begin{eqnarray}
C 
&=& 
-4 (q^A q^C)
\left( [p^B,p^A] [p^B,p^C]+ [q^B,p^A] [q^B,p^C] \right) 
\CR
&& -4 (p^A p^C)
\left( [q^B,q^A] [q^B,q^C]+ [p^B,q^A] [p^B,q^C] \right) 
\CR
&& +8 (p^A q^C)
\left( [q^B,q^A] [q^B,p^C] \right)
+8 (p^C q^A)
\left( [p^B,p^A] [p^B,q^C] \right)
+ {\cal O} ([\, ,]^3) , 
\label{123}
\end{eqnarray}
This term is not $SO(8)$ invariant, but $SU(4) \times U(1)$ invariant.
Finally, by substituting (\ref{real_im}) to $D$ defined in (\ref{defs}),
we obtain 
\begin{eqnarray}
D
&=& 
- 2 (p^A p^C +q^A q^C) 
\left( [p^B,p^A] [q^B,q^C]+ [q^B,q^A] [p^B,p^C]- [q^B,p^A] [p^B,q^C]
\right.
\CR
&& \qquad \qquad \qquad \qquad \qquad \left.
 - [p^B,q^A] [q^B,p^C] 
+2  [q^C,p^A] [p^B,q^B] \right)
\CR
&& + 2 (p^A q^C - p^C q^A)
\left( 2 [p^B,p^A] [q^B,p^C] - 2 [q^B,q^A] [p^B,q^C] 
\right.
\CR
&& \qquad \qquad \qquad \qquad \qquad \left.
+ \left( [p^C,p^A] +[q^C,q^A] \right) [p^B,q^B] \right)
+ {\cal O} ([\, ,]^3) .
\end{eqnarray}
This is also $SU(4) \times U(1)$ invariant.
By using the identity (\ref{id_xyz}) and the Jacobi identity,
similarly to the calculation in (\ref{B0}),
this can be rewritten as
\begin{eqnarray}
D
&=& 4 q^C \left( [p^B,p^2] [q^B,p^C]- [p^B,p^C] [q^B,p^2] 
     +[p^C,p^2] [p^B,q^B]\right)
\CR
&& + 4 p^C \left( -[p^B,q^2] [q^B,q^C]+ [p^B,q^C] [q^B,q^2] 
    - [q^C,q^2] [p^B,q^B] \right)
\CR
&& + {\rm total \, div.} + {\cal O} ([\, ,]^3) .
 \label{0124}
\end{eqnarray}

From (\ref{bos3}), (\ref{A}), (\ref{123}), and
(\ref{0124}), we have shown that the bosonic potential
is rewritten as
\begin{eqnarray}
 V_{\rm bos} &=& 
- \frac{4\pi^2}{3k^2} \mathrm{Tr}
\left(
6 ( (p^A)^2 + (q^A)^2) \left([p^B,p^C]^2 +2 [p^B,q^C]^2+
[q^B,q^C]^2 \right)  \right. \CR
&& -12 (q^A q^C)
\left( [p^B,p^A] [p^B,p^C]+ [q^B,p^A] [q^B,p^C] \right)
\CR
&& -12 (p^A p^C)
\left( [q^B,q^A] [q^B,q^C]+ [p^B,q^A] [p^B,q^C] \right) 
\CR
&& +24 (p^A q^C)
\left( [q^B,q^A] [q^B,p^C] \right)
+24 (p^C q^A)
\left( [p^B,p^A] [p^B,q^C] \right) \CR
&& + 12 q^C \left( [p^B,p^2] [q^B,p^C]- [p^B,p^C] [q^B,p^2] 
+[p^C,p^2] [p^B,q^B]\right)\CR
&& \left. 
+12 p^C \left( -[p^B,q^2] [q^B,q^C]+ [p^B,q^C] [q^B,q^2]-  [q^C,q^2] [p^B,q^B]
 \right)
\right)
\CR 
&&
+{\rm total \, div.} 
+ {\cal O} ([,]^3) .
\label{bos4}
\end{eqnarray}

\subsection{Expansion of the ABJM action}

In order to obtain an action for the M5-branes,
we will consider the fluctuations around 
the explicit classical solution which is obtained for 
$\Theta \rightarrow 0$ limit, where the terms with the least
numbers of the commutator should be kept.
Thus, we should impose how large the fluctuations are
compared with $\Theta$.

First, we  assume that the fluctuations of the scalar field is of
 ${\cal O} (\Theta^{\frac{1}{2}})$
and then the commutator of them is ${\cal O} (\Theta)$.
Because the backgrounds, $r$ and $r'$, are ${\cal O} (1)$,
the commutators between the backgrounds are ${\cal O} (\Theta)$.
If we compactify the theory on a circle,
the M5-brane effective action should be reduced to 
the D4-brane effective action.
Thus, we would like to have 
the Yang-Mills terms and kinetic terms for the scalars
kept in the action.\footnote{
To have a conformal M5-brane action, we may have to take
another assumption. In this paper, we take this assumption
in order to compare the result with the D4-brane action,
as a step toward finding the multiple M5-brane action.} 
As we will see later, 
the assumption that the fluctuations of scalar fields are
${\cal O} (\Theta^{\frac{1}{2}})$ is consistent with this.

Next, we consider the gauge fields
and the derivatives.
Introducing
\beq
\hat{z}^1=\hat{x}, \,\, \hat{z}^2=\hat{y}, \,\, 
\eeq
we have
\beq
[\hat{z}^a, \hat{z}^b] = i \Theta \epsilon^{ab},
\label{def_z}
\eeq
where $a,b=1,2$ and $\epsilon^{12}=\epsilon_{12}=1$.
In the standard procedure for the construction 
of the non-commutative D-brane from the matrix model 
\cite{BSS, CoDoSc, AIIKKT, Se},
the fluctuations  around the D4-brane solution
in the infinitely many D2-branes
are introduced as
\beq
\hat{X}^a=\
i \Theta \epsilon^{ab} \hat{D}_b
=\hat{z}^a- \epsilon^{ab} A_b, \;\;\;\;
\hat{D_a} \equiv i \epsilon_{ab} \hat{z}^b \Theta^{-1} + i A_a,
\label{def_capZ0}
\eeq
where $\hat{X}^a$ is the scalar fields of D2-branes
and $A_a$ is the fluctuations around 
the solution $\hat{X}^a=\hat{z}^a$.
{}From the definition of the covariant 
derivative operator $\hat{D}_a$, 
we have 
\beqa
[\hat{D}_a, f (\hat{z} ) ] &=& \partial_a f(\hat{z})+ i
[A_a,f(\hat{z})], \CR
{[} \hat{D}_a,\hat{D}_b {]}  &=& 
i \left( F_{ab} -\epsilon_{ab} \Theta^{-1} \right), \,\,\,\,
F_{ab} = \partial_a A_b -\partial_b A_a +i [A_a,A_b].
\eeq
In our case, imitating this we take the following
parametrization of the fluctuations:
\beq
\hat{Z}^a=\hat{z}^a+ fluctuation=
i \Theta \epsilon^{ab} \hat{D}_b + i \Phi^a, \;\;\;\;
\hat{D_a} \equiv i \epsilon_{ab} \hat{z}^b \Theta^{-1} + i A_a .
\label{def_capZ}
\eeq
Taking account that the scalar fields $Y_A$ in ABJM action are complex,
we have introduced the fluctuations $\Phi^a$ here for later convenience.
Since the fluctuations of the scalar fields are 
assumed to be ${\cal O} (\Theta^{\frac{1}{2}})$,
the covariant derivative operator 
$\hat{D}_a$ is ${\cal O} (\Theta^{-\frac{1}{2}})$,
which also means that both the derivative $\partial_b$ and the gauge fields
are ${\cal O} (\Theta^{-\frac{1}{2}})$.
On the other hand, we assume that the gauge field $A_{\mu}$, 
and the derivative $\partial_{\mu}$ $(\mu =0,1,2)$
for the three dimensional spacetime,
which are originally included in the ABJM action (\ref{ABJMaction2}),
are of order ${\cal O} (\Theta^{\frac{1}{2}})$.\footnote{
This means that we only consider 
the fluctuations $\Phi, A$ such that 
$\partial_\mu \Phi, \partial_\mu A$ are small compared 
with $\Phi, A$ by ${\cal O} (\Theta^{\frac{1}{2}} ) $, respectively.}
We will also assume that $B_{\mu} = {\cal O} (\Theta)$ 
$(\mu =0,1,2)$ which is consistent with the equations of motion
as will be seen later.
In this approximation, all the kinetic terms of D4-brane actions are kept
and the action will be ${\cal O} (\Theta)$.
We could have regarded the scale of the 
the each of the fluctuations are independent.
Here, we take the simplest and consistent one.

The approximated potential (\ref{bos4}) has been obtained 
by expanding the potential (\ref{bos})
up to two commutators and by 
substituting the decomposition (\ref{real_im}).
However, by assuming that the fluctuation of the scalar field is
 ${\cal O} (\Theta^{\frac{1}{2}})$, as stated previously,
we can also regard this as the expansion in 
the non-commutative parameter $\Theta$.
Then, we notice that only the terms of the classical solution
contribute to the factor outside of commutators 
while both the classical solution and the fluctuation 
contribute inside the commutators in (\ref{bos4}).
Although our classical solution (\ref{M5_solution}) has trivial
ambiguities, described in \cite{previous},
related to the area preserving diffeomorphism, 
we choose its explicit form as in (\ref{ansatz1}) with (\ref{ansatz2}),
where $q^A$ is zero for the classical solution.
Then, in this approximation, the potential reduces to 
\begin{eqnarray}
 V_{\rm bos} &=& 
- \frac{4\pi^2}{3k^2} \mathrm{Tr}
\left(
6  (p^A)^2  \left( [p^B,p^C]^2 +2 [p^B,q^C]^2+
[q^B,q^C]^2 \right)  \right. \CR
&& \qquad \qquad
\left. 
-12 (p^A p^C)
\left( [q^B,q^A] [q^B,q^C]+ [p^B,q^A] [p^B,q^C] \right) 
\right)
\CR
&&
+ {\rm total \, div.} + {\cal O} (\Theta ^{5/2}),
\label{rpot}
\end{eqnarray}
where the quadratic terms of $p$ outside the commutators,
which will be approximated by the classical value, 
are remained.


Next, we will consider the all of the bosonic part of the action, 
i.e. including the Chern-Simons term and the kinetic terms.
The covariant derivative (\ref{cov}) is rewritten
in terms of $p$ and $q$ in (\ref{real_im}) as 
\begin{eqnarray}
\tilde{D}_{\mu} Y^A 
&=& D_{\mu} p^A -\{ B_{\mu}, q^A \} +i \left(
D_{\mu} q^A +\{ B_{\mu}, p^A \} \right).
\label{cov2}
\end{eqnarray}
The equations of motion for $B_\mu$ is
obtained from (\ref{ABJMaction2}) and (\ref{cov2}) as
\beq
0 = \frac{k}{2 \pi} \epsilon^{\mu \nu \rho} F_{\nu \rho}
+4 q^A \left( D_{\mu} p^A -\{ B_{\mu}, q^A \} \right)
-4 p^A \left( D_{\mu} q^A +\{ B_{\mu}, p^A \} \right)
+\frac{2 i k}{2 \pi} \epsilon^{\mu \nu \rho} B_\nu B_\rho.
\eeq
Keeping the leading order terms in $\Theta$,
we can solve this equations of motion as
\beq
B_\mu \simeq \frac{1}{2 (p_A)^2} 
\left( -p^B D_\mu q^B + \frac{k}{8 \pi} 
\epsilon^{\mu \nu \rho} F_{\nu \rho} \right),
\label{solB}
\eeq
where factor $(p^A)^2$ in the denominator in (\ref{solB}) as well as that appears
in the following are always evaluated as its classical value in our approximation.
The solution (\ref{solB}) is consistent with the assumption that $B_{\mu} = {\cal O}
(\Theta)$,
which we have imposed above.
Because the Lagrangian is quadratic in $B_\mu$ in the approximation,
we can integrate it out to obtain
\begin{eqnarray}
L_{bos} \simeq 
- {\rm Tr} \left( 
\left( D_{\mu} p^A \right)^2+
\left( D^{\mu} q^A \right)^2 
-\frac{\left(p^B D_\mu q^B \right)^2}{(p_A)^2}
+\frac{k^2}{16 \pi^2 (p_A)^2} F^2
\right)
- V_{\rm bos},
\end{eqnarray}
where we denoted $F^2 \equiv F_{\mu\nu} F^{\mu\nu}$.
Therefore in the approximation $\Theta \to 0$, we have 
\begin{eqnarray}
L_{bos} &\simeq& 
- {\rm Tr} \left( 
\left( D_{\mu} p^A \right)^2+
\left( D^{\mu} q^A \right)^2 
-\frac{\left(p^B D_\mu q^B \right)^2}{(p_A)^2}
+\frac{k^2}{16 \pi^2 (p_A)^2} F^2
\right. \CR
&& -\frac{8\pi^2}{k^2} 
  (p^A)^2  \left( [p^A,p^B]^2 +2 [p^A,q^B]^2+
[q^A,q^B]^2 \right)   \CR
&& 
\left.
+\frac{16\pi^2}{k^2}  (p^A p^C)
\left( [q^B,q^A] [q^B,q^C]+ [p^B,q^A] [p^B,q^C] \right) 
+{\rm total \, div.} \right).
\label{app_action}
\end{eqnarray}

\subsection{Action of the fluctuations around the solution}

Now we evaluate the action of the fluctuations explicitly.
For $A=3,4$, we set
\beq
Y^A=p^A+i q^A= \Phi^{2 A-3} + i \Phi^{2 A-2},
\eeq
where $\Phi^3, \Phi^4, \Phi^5, \Phi^6$ are Hermite operators.
Here, we represent our M5-branes solution as
\beq
Y^A=Y^A(\hat{z}^b),
\eeq
where $A=1,2$ and $\hat{z}^b$ satisfies (\ref{def_z}).
Then, the fluctuations around it are introduced by
\beq
Y^A=Y^A(\hat{Z}^b)
\eeq
with $\hat{Z}^b$ defined as in (\ref{def_capZ}),
where we keep the orderings of $\hat{z}$s and $\hat{Z}$s.
In the Poisson bracket approximation,
$Y^a(\hat{z})$ are Hermite and
\beq
J \epsilon^{AB} \equiv [ Y^A, Y^B]_P =
\frac{i \Theta v}{\sqrt{(Y^A)^2}} \epsilon^{AB}.
\eeq
In the commutator, the scalar fields can be replaced by
\begin{eqnarray}
 Y^A 
&\simeq& 
i \Theta \frac{\de Y^A}{\de z^c} \epsilon^{cb} \hat{D}_b 
+ i \frac{\de Y^A}{\de z^b} \Phi^b \CR
&=& 
J \epsilon^{AB} D_B + i \Phi^A,
\end{eqnarray}
where we introduced 
\beq
D_B \equiv \frac{\de z^b}{\de Y^B} \hat{D}_b, \;\;\;
\Phi^A \equiv \frac{\de Y^A}{\de z^b} \Phi^b.
\eeq
Note that in the approximation, 
$D_B$ act as the derivative with respect to $Y^B$ in the commutator
because $\frac{\de Y^B(Z)}{\de z^b} \sim \frac{\de Y^B(z)}{\de z^b} 
{\rm + fluctuations}$, 
and then 
$[D_B, f] = \frac{\de z^b}{\de Y^B} [\hat{D}_b, f]+
[\frac{\de z^b}{\de Y^B},f] \hat{D}_b \simeq 
\frac{\de z^b}{\de Y^B} [\hat{D}_b, f]$.
Thus
\beq
p^A  \simeq J \epsilon^{AB} D_B, \,\,
q^A \simeq  \Phi^A. 
\eeq

Now we will rewrite the approximated action (\ref{app_action}) 
with the above terms.
The first line includes the kinetic terms for $p,q$ 
in the direction of original three dimensional spacetime which M2 branes extend.
However, the kinetic term for the scalar field defined as
\begin{eqnarray}
 \Phi_{\|} \equiv \frac{1}{(p^A)^2} p^B q^B
\end{eqnarray}
is subtracted.
The fields $p^A$ and $p^B$ should 
be regarded as the classical value in our approximation.
This $\Phi_{\|}$ is the fluctuation for the direction
generated by the $U(1)_b$ gauge symmetry of the ABJM action
from the classical value and more explicitly\footnote{
We use the representation with the star-product.}
\beq
 \Phi_{\|} \sim \frac{1}{\sqrt{r^2+{r'^2}}}
(r \Phi^1 + r' \Phi^2).
\eeq
Denoting the orthogonal part of the scalar fields as, 
\beq
\Phi_{\bot}^i \equiv \left\{  \frac{1}{\sqrt{r^2+{r'^2}}}
(r' \Phi^1 - r \Phi^2),\Phi^3,\Phi^4\,\Phi^5,\Phi^6 \right\},
\label{Phi_orth}
\eeq
we find that the first line in (\ref{app_action}) 
includes kinetic terms for $\Phi_{\bot}^i$ as well as those 
for $p^A$ ($A=1,2$).
The kinetic terms for $p^A$ ($A=1,2$)
can be rewritten as $J \varepsilon^{AB} [D_{\mu},D_{B}]$.
Thus, first line in (\ref{rpot}) is given as
\begin{eqnarray}
\sim (D_{\mu} \Phi_{\bot}^i) ^2 - 2 J^2 [D_{\mu},D_{B}]
+\frac{k^2}{16 \pi^2 (p_A)^2} F^{\mu\nu} F_{\mu\nu} .
\label{first}
\end{eqnarray}
The second line in (\ref{app_action}) is straightforwardly  
shown to give Yang-Mill like terms 
\beq
\sim 6 (r^2+{r'^2}) 
([\Phi^i,\Phi^j]^2 - 2 J^2 [D_B,\Phi^i]^2+ J^4 [D_B,D_C]^2),
\label{second}
\eeq
where $i,j=1\ldots6$.
The third line can be also rewritten in terms of 
$\Phi_{\|} $ by rewriting $p_A q_A \sim (r^2+r'^2) \Phi_{\|}$ as 
\begin{eqnarray}
\sim 12 (r^2+{r'^2}) 
([\Phi^i,\Phi_{\|}]^2 - 2 J^2 [D_B,\Phi_{\|}]^2).
\end{eqnarray}
Again, this term subtract the contribution
of $\Phi_{\|}$ from (\ref{second})
and remaining terms are those for ${\Phi_{\bot}^i}$.
Thus, the scalar field $ \Phi_{\|}$ completely disappears. 
This is a consequence of the Higgs mechanism described in \cite{Mukhi}.


Adding all the contributions above, we obtain 
\begin{eqnarray}
 L_{\rm bos} &\simeq& 
-\mathrm{Tr}
\left(
 \frac{4\pi^2}{3k^2} 
6  
(r^2+{r'^2}) 
\left(
[\Phi^i_{\bot},\Phi^j_{\bot}]^2 - 2 J^2 [D_B,\Phi^i_{\bot}]^2
+ J^4 [D_B,D_C]^2
\right) \right. \CR
&& \left.
+(D_\mu \Phi_{\bot})^2 - 2 J^2 [D_{\mu}, D_{A} ]^2 
+\frac{k^2}{16 \pi^2 (r^2+{r'}^2)} F_{\mu \nu} F^{\mu \nu}
\right)
+{\rm total \, div.},
\label{rpot3}
\end{eqnarray}
where 
\beq
J^2= -\frac{\Theta^2 v^2}{r^2+{r'^2}}.
\eeq
The trace can also be replaced by 
\beq
 \mathrm{Tr} \rightarrow 
\int dr dr' \frac{\sqrt{r^2+{r'^2}}}{2 \pi \Theta v}.
\eeq

To rewrite this simpler,
we further introduce the analogue of the 
open string metric \cite{SW} as
\beq
g_{rr}=g_{r' r'}=\frac{k^2}{16 \pi^2 \Theta^2 v^2}
=8 \pi^2 H^2,
\eeq
where $H$ is the constant flux on the M5-branes 
\beq
H \equiv i F_{012} =\frac{k}{8 \sqrt{2}  \pi^2 \Theta v},
\eeq
and the index $M=\{ \mu,A \}$ which runs the directions 
of longitudinal to the D4-branes.
Then up to the total divergence, we have
\begin{eqnarray}
S_{\rm bos} & \simeq& const. + \int d^3 x dr dr' 
\frac{1}{2 ( g_{\rm YM})^2} 
\left[  -8 \mathrm{tr} (F^{MN} F_{MN})
 - 2 {\rm tr} (D_{M} \phi_{\bot}^i D^{M} \phi_{\bot}^i)
-{\rm tr} \left( [ \phi_{\bot}^i,  \phi_{\bot}^j ]^2
\right) 
\right], \CR
\label{af}
\end{eqnarray}
where
\beq
\phi_{\bot}^i \equiv  g_{YM} \frac{\sqrt{r^2+{r'}^2} }{2 \pi \Theta v} 
\Phi_{\bot}^i,
\eeq
and the (non-constant) 5-dimensional gauge coupling as
\begin{eqnarray}
\frac{1}{g_{\rm YM}^2} 
\equiv  \frac{k^2}{16 \pi^3 \Theta v \sqrt{r^2 + r'^2}}
=\frac{k H}{\sqrt{2} \pi \sqrt{r^2+{r'}^2}}.
\label{5dg}
\end{eqnarray}
The constant term was already computed in \cite{previous}
and gives the correct tension of the M5-branes.
This action is considered as the 
the action of D4-branes with non-constant dilaton 
background.
Indeed, the $r$ and $r'$ dependence of the gauge coupling 
is correct one.
For an M5-brane, the action is consistent with
the known one if we take into account the fact 
that we keep only the zero-mode of the 
$\theta$-direction and the action can be dimensionally 
reduced to 5-dimension.
For the multiple M5-branes, if we drop the 
non-zero modes, 
we expect the action will be the action of the D4-branes
with the gauge coupling (\ref{5dg}),
(bosonic part of) which is the action (\ref{af}).

As discussed in section \ref{Intro},
the action with the non-constant gauge coupling
is not obtained from the D2-brane action
and our result here is non-trivial.
Of course, the really interesting problem 
is to include the non-zero modes of the $\theta$-direction
by considering the monopole operators.
We hope our result will be an useful 
for investigating it.

\section{Discussion}

In this paper, we have calculated the fluctuation 
from the classical M5-brane solution of ABJM model
and obtained the action for D4-branes with non-constant dilaton background.
In order to understand the low energy dynamics
of multiple M5-brane dynamics more in detail,
we mention several points 
which we should improve in our analysis.

First, in this paper, we have ignored the 
the total divergence term, which include
the terms vanish by taking the trace naively.
However, such terms should be important
and correspond to topological terms.
Indeed, in the construction of the usual D3-branes \cite{HTT}
from the orbifolded ABJM action \cite{Be}-\cite{TeYa}, 
such term gives the correct $\theta$-term
on the D3-branes.

Second, we should include the contribution
from the monopole operators,
which we have already discussed above
for supplementing the KK modes of the gauged $U(1)$ direction.
This problem will be related to the very recent argument 
that the KK modes will be
present in 5D super Yang-Mills theory \cite{Do, LaPaSc}.
The singularity of $\mathbb{C}^4/ \mathbb{Z}_k$ 
might be important and should be carefully considered.

Finally, it is interesting to extend our analysis to the case 
of M5-branes with finite magnetic flux.
In our analysis, the commutative limit is considered,
which corresponds to the limit that the magnetic flux is infinitely strong.
Since the classical solution for the finite non-commutative parameter 
is known only approximately \cite{previous}, 
we also need to develop this point.

We hope to do more careful analysis in order 
to understand these points in near future.


\section*{Acknowledgments}
We would like to thank 
K. Hosomichi, Y. Imamura, S. Sugimoto, and Piljin Yi 
for useful discussions.
S.~T.~is partly supported by the Japan Ministry of Education, Culture,
Sports, Science and Technology. 
The work of F.~Y.~is partly supported by the William Hodge Fellowship.

\end{document}